# Functional Architecture of the Human Hypothalamus: Cortical Coupling and Subregional Organization Using 7-Tesla fMRI

**Authors**

Kent M. Lee[1]

Joshua Rodriguez[1]

Ludger Hartley[1]

Jiahe Zhang[1,2,3]

Philip A. Kragel[4]

Lorena Chanes[5]

Tor D. Wager[6]

Karen S. Quigley[1]

Lawrence L. Wald[7]

Marta Bianciardi[7]

Lisa Feldman Barrett[1,2,7]

Jordan E. Theriault[1]

Ajay B. Satpute[1,7]

Affiliations:

[1]Department of Psychology, Northeastern University, Boston, MA

[2]Department of Psychiatry, Massachusetts General Hospital, Boston, MA

[3]Department of Psychiatry, Harvard Medical School, Boston, MA

[4]Department of Psychology, Emory University, Atlanta, GA

[5]Department of Clinical and Health Psychology, Universitat Autònoma de Barcelona, Barcelona, Spain

[6]Department of Psychological and Brain Sciences, Dartmouth College, Hanover, NH

[7]Department of Radiology, Massachusetts General Hospital, Boston, MA

**Abstract**

The hypothalamus plays an important role in the regulation of the body's metabolic state and behaviors related to survival. Despite its importance however, many questions exist regarding the intrinsic and extrinsic connections of the hypothalamus in humans, especially its relationship with the cerebral cortex. As a heterogeneous structure, it is possible that the hypothalamus is composed of different subregions, which have their own distinct relationships with the cerebral cortex. Previous work on functional connectivity in the human hypothalamus have either treated it as a unitary structure or relied on methodological approaches that are limited in modeling its intrinsic functional architecture. Here, we used resting-state data from ultra-high field 7-Tesla fMRI and a data-driven analytical approach to identify functional subregions of the human hypothalamus. Our approach identified four functional hypothalamic subregions based on intrinsic functional connectivity, which in turn showed distinct patterns of functional connectivity with the cerebral cortex. Overall, all hypothalamic subregions showed stronger connectivity with a cortical network (Cortical Network 1) composed primarily of frontal, midline, and limbic cortical areas and weaker connectivity with a second cortical network composed largely of posterior sensorimotor regions (Cortical Network 2). Of the hypothalamic subregions, the anterior hypothalamus showed the strongest connection to Cortical Network 1, while a more ventral subregion containing the anterior hypothalamus extending to the tuberal region showed the weakest connectivity. The findings support the use of ultra-high field, high-resolution imaging in providing a more incisive investigation of the human hypothalamus that respects its complex internal structure and extrinsic functional architecture.

## Introduction

The hypothalamus lies at the interface between neural and endocrine systems, enabling central nervous system regulation over autonomic bodily functions. As such, it plays a critical role in the regulation of the body's internal state and behaviors related to survival (Lechan & Toni, 2016; Saper, 2012; Saper & Lowell, 2014). To serve these functions, the hypothalamus has complex networks of internal and external connections. Internally, the hypothalamus is composed of a collection of diverse nuclei that are implicated in diverse functions including regulating circadian rhythms, metabolism, feeding, body temperature, and sexual and reproductive behavior (Burbridge et al., 2016; Lechan & Toni, 2016; Saper, 2012; Saper & Lowell, 2014). Externally, hypothalamic subregions have a widespread pattern of connections with numerous cortical areas that are important for external sensory integration and processing viscerosensory information (Barbas, 2000; Çavdar et al., 2001; Freedman et al., 2000; Kleckner et al., 2017; Öngür et al., 1998; Rempel-Clower & Barbas, 1998). The neuroanatomical connections of the hypothalamus, and its complex internal organization enable it to serve its critical role in metabolic regulation (Burbridge et al., 2016; Lechan & Toni, 2016; Luiten et al., 1987; Qu et al., 1996; Saper, 2012; Saper & Lowell, 2014; Wirth et al., 2001; Yokosuka et al., 1999).

Despite its central role in survival, many unanswered questions remain about the functional architecture of the human hypothalamus. Namely, these questions concern the relationships between subregions of the hypothalamus and their connections to the cerebral cortex. One critical limitation is that there is a paucity of work on the hypothalamus of primates in general, human or otherwise (Saper, 2012). Only limited postmortem work exists of the human hypothalamus (e.g., Abrahamson et al., 2001; Dai, Swaab, et al., 1998; Dai, Van Der Vliet, et al., 1998; de Lacalle & Saper, 2000; Panula et al., 1990; see Saper, 2012) and this body of work faces limitations in the scope of the studies and in sample size. By their nature, these postmortem studies also cannot tell us about the functional connections between the hypothalamus and the cerebral cortex. Thus, much of the received wisdom about the human hypothalamus relies on extensions from work on non-human vertebrates. However, as we discuss below, significant cross-species differences complicate attempts to directly extrapolate this non-human animal work to humans. These are limitations our ultra-high field 7-Tesla (T) fMRI approach aims to overcome.

**Cross-species studies of the hypothalamus**

The hypothalamus has been studied across a diversity of non-human animals, including zebrafish, rodents, bats, cats, sheep, and some limited non-human primates (Burbridge et al., 2016; Henry, 2003; King & Anthony, 1984; Löhr & Hammerschmidt, 2011; Markakis, 2002; Saper, 2012; Saper et al., 1978; Xie & Dorsky, 2017). In turn, many assumptions about the human hypothalamus have been extrapolated from this body of work. However, although studies of comparative anatomy have revealed many commonalities between vertebrate hypothalami (Markakis, 2002; Saper, 2012; Xie & Dorsky, 2017), they have also found important cross-species differences in which the structure and connections of hypothalami can vary substantially, even across vertebrates (e.g., Chometton et al., 2016; King & Anthony, 1984; Panula et al., 1990). Two notable differences have to do with the cytoarchitecture and orientation of the human hypothalamus. First, compared to other mammals, nuclei in the human hypothalamus have less well-defined boundaries between one another (Dudás, 2021; Lechan & Toni, 2016; Saper, 2012). Second, the orientation of the human hypothalamus also differs compared to other mammals because the human skull is proportionally shorter in the anterior-posterior direction. In comparative anatomical studies, slices of parts of the hypothalamus are taken but due to the orientation of the human hypothalamus, the angle of the slices can vary up to 45-degrees, resulting in homologues of the human hypothalamus being anteroflexed, retroflexed, or even vertical compared to other animals (Saper, 2012). Further, there are differences in the exact locations of hypothalamic cell populations, cell morphology, and projections (Bresson et al., 1989; Croizier et al., 2011; Dai, Van Der Vliet, et al., 1998; Johnson et al., 1988; King & Anthony, 1984; Levine et al., 1991; Swanson et al., 2005), which are compounded by variation in orientation, and that collectively pose an additional challenge to identifying analogous subregions within the human hypothalamus based on their locations in non-human animals.

**Using fMRI to investigate the intra- and interconnections of human hypothalamus with cortical areas**

Functional magnetic resonance imaging (fMRI) provides unique opportunities to address gaps in knowledge about the human hypothalamus and its relationship to the cerebral cortex in vivo. Specifically, an important question is whether the human hypothalamus acts as a unitary structure or is composed of different functional subunits with their own relationships to cortical areas. There is extensive non-human animal work showing that the hypothalamus is composed of a heterogeneous collection of nuclei (Lechan & Toni, 2016; Saper, 2012; Saper & Lowell, 2014). Yet, two important questions are whether fMRI can provide sufficient resolution to identify subregions within small subcortical structures such as the hypothalamus, and whether anatomical boundaries in non-human animals correspond to functional subregions, if any, of the hypothalamus in humans.

The majority of the existing functional connectivity studies on the human hypothalamus have relied on 3-Tesla, or weaker, field strength (e.g., Gao et al., 2020; Hinkle et al., 2013; Hirose et al., 2016; Kullmann et al., 2014; Lukoshe et al., 2017; Moulton et al., 2014; Qiu et al., 2013; Wright et al., 2016). These studies have also typically treated the hypothalamus as a unitary functional structure, and focused on comparisons of hypothalamic activity between different states (e.g., during glucose or leptin replenishment; Hinkle et al., 2013; Wright et al., 2016 or in vs. out of migraine attack; Moulton et al., 2014) or between healthy and patient populations (e.g., chronic premature ejaculation patients; Gao et al., 2020; migraine and cluster headache patients; Moulton et al., 2014; Qiu et al., 2013; Prader-Willi syndrome; Lukoshe et al., 2017).

To our knowledge, only two studies have examined functional connectivity in hypothalamic subregions in neurotypical samples (Hirose et al., 2016; Kullmann et al., 2014). In both studies, the researchers divided the hypothalamus into lateral and medial subregions using a seed-based approach with 3T fMRI data during resting-state activity, or activity while participants are awake but not engaged in a specific task or activity. Kullman et al. (2014) found that the lateral hypothalamus showed greater functional connectivity with midline anterior cingulate cortex (ACC), right insula, and right angular gyrus. Conversely, the medial hypothalamus showed greater functional connectivity with the superior orbitofrontal cortex (OFC), superior frontal gyrus (SFG), precuneus, cuneus, and middle occipital gyrus. Hirose et al. (2016) only examined the differences between lateral and medial hypothalamic functional connectivity with the OFC. They found that the lateral hypothalamus showed stronger connectivity with the medial OFC, while the medial hypothalamus showed stronger connectivity with the lateral OFC. These studies are suggestive of the notion that fMRI is capable of identifying functional subregions of the human hypothalamus. Yet, the seed-based approach used in these studies relied on placing spheres surrounding a particular coordinate location, or regions of interest assumed on the basis of anatomical boundaries from prior work. It is unclear whether these gross anatomical subdivisions of the hypothalamus are actually justified by the fMRI data itself (i.e., using data-driven techniques), and further, whether more finer grained parcellations could be achieved using more advanced fMRI techniques.

**The Current Study**

In order to overcome limitations of prior studies and identify any functional subregions within the hypothalamus, we used ultra-high field strength, high resolution fMRI at 7-Tesla to generate the first, data-driven, functional parcellation of the human hypothalamus. When combined with 32-channel head coils, 7T fMRI provides improvements in signal-to-noise ratio nearly an order-of-magnitude higher than 3T methods (Keil et al., 2010) used in  prior studies (e.g., Hirose et al., 2016; Kullmann et al., 2014). 7T fMRI also provides increased sensitivity to microvasculature (Duyn, 2012). In our prior work, this technique enabled us to obtain fMRI data at 1.1 mm isotropic voxel resolution (compared to the 3 mm isotropic resolution in prior work) and identify functional subregions even within small, subcortical structures, such as such as the periaqueductal gray (Fischbach et al., 2024; Kragel et al., 2019; Satpute et al., 2013) and superior colliculi (Chen, Kragel, Savoca, et al., 2022; Chen, Kragel, Wager, et al., 2022; Kragel et al., 2021; Wang et al., 2020). Here, we combine 7T fMRI with a data-driven analytical approach to identify functional subregions in the hypothalamus with resting fMRI data and their respective functional connectivity with cortical structures throughout the brain. In doing so, we provide the first, data-driven functional connectomic map of human hypothalamic subregions.

## Results

### Functional Subregions of the Hypothalamus

To examine functional subregions of the hypothalamus, we analyzed resting-state data using 7T fMRI from 104 adult participants. We first performed Louvain community detection on the group level functional connectivity matrix, which assigns each voxel to a community. Next, to ensure robustness of our results, we iterated the Louvain community detection 100 times, calculated the proportion of times each pair of voxels were assigned to the same community, and then applied the Louvain community detection algorithm again to the proportion matrix. **Table 1** shows the mean probability for each community that a given pair of voxels within a community were assigned to the same community across 100 iterations of the Louvain algorithm. Using this approach, we identified four hypothalamic communities (see **Figure 1**). The first community included anteroventral portions of the anterior hypothalamus and extended to the tuberal hypothalamus (**Figure 1**, yellow). We will refer to this community as the Anteroventral-Tuberal Community. The second community included mostly anterior regions of the hypothalamus (**Figure 1**, red) and thus we will refer to it as the Anterior Community. The third community identified included areas primarily in the tuberal and posterior regions of the hypothalamus. In the superior-to-inferior plane, this community was located in the middle of the hypothalamus (**Figure 1**, green). We will refer to this community as the Middle Tuberal-Posterior Community. Finally, the fourth community included the superior parts of the hypothalamus and we thus call it the Superior Hypothalamic Community (**Figure 1**, blue).

**Cortical Connectivity with Hypothalamic Subclusters**

To examine the functional connectivity of these hypothalamic subregions with cortical areas, we created masks based on each subregion. We then extracted timeseries data for each community for each run and each subject. Timeseries data per ROI were concatenated across three functional runs per subject and intercorrelated. They were then Fisher z-transformed to generate a 364x364 (360 cortical ROIs and 4 hypothalamic subregion ROIs) functional connectivity matrix per subject. Group averaged Fisher z-transformed correlations are shown in **Figure 2**. To reduce multiple comparison concerns and simplify results, we performed a k-means cluster analysis to group together ROIs with similar connectivity profiles across hypothalamic subregions (i.e. across a 360 cortical ROIs x 4 hypothalamic subregion ROIs, matrix). Fisher z-values for ROIs in the same k-means cluster were averaged together for each subject. To identify the number of clusters that were justified by the data, we iteratively calculated the Calinski-Harabasz index across k-thresholds of 2 to 40, and we used the "kneedle" algorithm (Satopaa et al., 2011) to select an optimal solution at the elbow (see **Supplementary Figure 1**).

The first cortical network (orange in **Figure 3a**) contained primarily anterior and midline regions. Regions in Cortical Network 1 included the OFC, ventromedial and dorsomedial PFC, ventrolateral PFC, cingulate cortex, and precuneus. They also included an anterior portion of the superior temporal gyrus, the insula, and both medial and lateral portions of the superior frontal gyrus (SFG) and middle frontal gyrus (MFG). The second network (teal in **Figure 3a**) included mostly lateral and posterior brain regions, though it also included a more ventral part of the dorsomedial PFC. Other regions included in Cortical Network 2 were the dorsolateral PFC and broad swathes of sensorimotor areas, such as the pre- and postcentral gyri, temporal lobe, parietal lobe, and the occipital lobe.

We probed for differences between the cortical networks in their functional connectivity with the different hypothalamic communities using a 2 (cortical networks) x 4 (hypothalamic communities) repeated measures analysis of variance (ANOVA). We found a main effect of cortical network, $F$ (1, 103)

= 175.57, $p < .001$, showing that Cortical Network 1 had greater connectivity with the hypothalamic communities ($M = .12$, $SD = .05$) than Cortical Network 2 ($M = .03$, $SD = .03$). We also found a main effect of hypothalamic communities , $F$ (2.12, 217.99) = 29.60, $p < .001$, suggesting that there were differences between the communities in their connections with the cortical networks.

These main effects were qualified by a significant two-way interaction, $F$ (2.24, 230.80) = 33.28, $p < .001$. Post-hoc Least Significant Difference (LSD) tests revealed that the interaction was driven by differences in the strength of connectivity to each Cortical Network across the hypothalamic communities. Cortical Network 1 was significantly related to all hypothalamic communities. Cortical Network 2 had significant connectivity with all hypothalamic communities except the anteroventral-tuberal hypothalamic communities. We also conducted post hoc LSD tests comparing the connectivity of each hypothalamic community with each cortical network. The hypothalamic community most strongly connected with Cortical Network 1 ($M_{diff} \geq .04$, $p$'s < .001) and Cortical Network 2 ($M_{diff} \geq .06$, $p$'s < .001) was the anterior hypothalamic community in pairwise comparisons of the functional connectivity between each hypothalamic community and each cortical network (see **Figure 3b**).

## Discussion

We identified four, data-driven functional communities within the human hypothalamus based on intrinsic connectivity (see **Figure 1**) using an ultra-high field strength, high resolution 7T resting fMRI protocol. The communities can be broadly divided into anteroventral-tuberal, anterior, middle tuberal-posterior, and superior hypothalamic communities. All four communities showed overall greater functional connectivity with a network of cortical areas spanning anterior medial PFC, cingulate cortex, superior frontal gyrus, and lateral orbitofrontal cortex (see **Figure 3a**)—many of which have structural connections with the hypothalamus (Öngür et al., 1998; Rempel-Clower & Barbas, 1998; Risold et al., 1997). This relationship was stronger for the anterior community, followed by the superior community, relative to other hypothalamic communities (see **Figure 3b**). All four communities also showed relatively lower functional connectivity with a second network of largely posterior cortical areas and dorsolateral prefrontal cortex (see **Figures 2** & **3**). Yet here, too, the anterior community continued to exhibit stronger functional connectivity with these cortical areas relative to the other hypothalamic communities (see **Figure 3b**). Critically, a significant interaction indicated that the relationship between hypothalamic community to cortical connectivity depended on the hypothalamic subregion and cortical network, highlighting both the particularly strong relationship of the anterior community, and weak relationship of the anteroventral-tuberal community, with these cortical networks. These results suggest that this data-driven approach can be used to identify and investigate functional subregions within the human hypothalamus, and further, provide a first look at the structure of their functional relationships with cortical areas in humans.

**Intrinsic Connectivity Captures Functional Subregions of the Human Hypothalamus**

The hypothalamus is a heterogeneous structure composed of distinct nuclei with specialized functions (Lechan & Toni, 2016; Saper, 2012; Saper & Lowell, 2014) and our results indicate that this heterogeneity is reflected in the intrinsic and extrinsic functional connections of the hypothalamus. Yet, due to limitations in scanning resolution and signal-to-noise ratio, most fMRI studies have treated the hypothalamus as a unitary structure (e.g., Gao et al., 2020; Hinkle et al., 2013; Lukoshe et al., 2017; Moulton et al., 2014; Qiu et al., 2013; Wright et al., 2016). A couple of studies have attempted to anatomically divide the hypothalamus based on gross spatial organizations — such as lateral versus medial (Hirose et al., 2016; Kullmann et al., 2014) or anterior and posterior subdivisions (e.g., Schulte et al., 2017; Smeets et al., 2005).

In contrast, our data-driven approach identifies functional communities based on intrinsic connectivity. These communities do not strictly align with the more gross models, but they do show greater alignment with other anatomical frameworks of hypothalamic organization in humans (Makris et al., 2013; see also Billot et al., 2020; Rodrigues et al., 2024). For example, we did not find a strong lateral-to-medial division despite its prominence in cytoarchitectural models (Lechan & Toni, 2016; Saper, 2012). This division is also not apparent in studies that define hypothalamic subregions using structural MRI data (Billot et al., 2020; Makris et al., 2013; Rodrigues et al., 2024). Indeed, these studies divided the hypothalamus into five subregions that show similar correspondence to the anteroventral-tuberal, anterior, middle-tuberal, and superior hypothalamic communities we observed here (see **Figure 4**). It may be that the intrinsic functional connectivity of the hypothalamus does not closely align with the underlying cytoarchitecture. It is also possible that this lack of lateral-to-medial distinction may also be explained by the less differentiated boundaries between nuclei in the human hypothalamus. The functionally derived communities we observed here may be used as seeds, perhaps in comparison with more gross anatomical models, to guide future work on the functional architecture of the human hypothalamus.

**Functional Connectivity between Hypothalamic Subregions and the Cerebral Cortex**

Our findings on functional connectivity between hypothalamus and the cerebral cortex are consistent with previous studies in non-human animals examining anatomical connections. We found that a network of lateral and medial prefrontal areas, the insula, and cingulate cortex had robust functional connectivity with all hypothalamic subregions (**Figure 3a**, orange). In rodents, direct cortical projections to the hypothalamus originate primarily from prefrontal areas, the insula, and other infralimbic regions (Risold et al., 1997; Saper, 2012). We also found that the anterior hypothalamic subregion, followed by the superior hypothalamic subregion, had greater functional connectivity with prefrontal and insular cortical regions, relative to the other hypothalamic subregions. Our results are consistent with prior anatomical work in which direct projections from the prefrontal and insular regions terminate primarily in the preoptic and anterior hypothalamic nuclei (Risold et al., 1997; Saper, 2012), which were likely contained in the anterior hypothalamic subregion. The prefrontal cortex and the insula also send direct projections to the dorsomedial nucleus (Risold et al., 1997; Saper, 2012), the dorsal portion of which was likely contained in the superior hypothalamic community. Meanwhile, the community that showed the weakest functional connectivity with the cerebral cortex was the anteroventral tuberal subregion. This area likely contains certain nuclei that are not known to have strong connections to these same cortical structures (e.g., supraoptic and arcuate nuclei; albeit this area may also include the ventral most parts of the dorsomedial nucleus). These findings suggest that the neuroanatomical connections between the cerebral cortex and subregions of the hypothalamus may be reflected in their functional connectivity as well.

**Hypothalamic Connectivity with the Default Mode Network and Allostatic Network**

Our results revealed that several of the prefrontal cortical areas showing greater functional connectivity with the hypothalamus are part of the default mode network—a set of brain areas that are typically more active during rest (i.e., when not engaged in certain types of cognitive tasks) and that are structurally and functionally interconnected with one another (Biswal & Uddin, 2025; Greicius et al., 2003). The default mode network has been associated with a wide variety of psychological states (Barrett & Satpute, 2013; Biswal & Uddin, 2025; Buckner & Carroll, 2007; Raichle, 2015) that in common involve greater levels of abstraction in processing sensory input (Barrett, 2017; Margulies et al., 2016; Satpute & Lindquist, 2019; Smallwood et al., 2021), including episodic memory (Ranganath & Ritchey, 2012; Sestieri et al., 2011), language (Fernandino & Binder, 2024), social cognition (Spunt et al., 2011), and emotion representation (Barrett, 2017; Lee & Satpute, 2024; Satpute & Lindquist, 2019).

Notably, the default mode network is not necessarily a monolithic network but comprises several networks and/or states (Andrews-Hanna et al., 2010; Buckner & DiNicola, 2019; Ciric et al., 2017; Ranganath & Ritchey, 2012). Recent cytoarchitectonic and structural connectivity findings suggest that more anterior nodes, including the anteromedial and ventrolateral prefrontal cortex, are more isolated from exteroceptive sensory input relative to more posterior nodes in the network (e.g., the precuneus and temporoparietal area; (Paquola et al., 2025). In our data, these anterior prefrontal regions exhibited stronger connectivity with the hypothalamus, suggesting that they may be preferentially involved in viscerosensory and visceromotor processes (Kleckner et al., 2017). This anterior subset also clustered with other regions known for their functional role in bodily regulation, including the anterior and posterior cingulate cortex and the insula, reinforcing the idea that functional differentiation within the default mode network may reflect differences in connection patterns with external v. internal or visceral sensory systems (Paquola et al., 2025). These findings concerning the default mode network also align with the recent formulation of an allostatic network—a network for predictive regulation of the body's internal state (Girn et al., 2024; Kleckner et al., 2017; Zhang et al., 2025; also see a "revised limbic network" model, Catani et al., 2013). By this account, this functional coupling may reflect the dynamic interaction of predictions driven by cortical areas of the allostatic network (particularly allocortex as well as agranular and dysgranular cortices) and prediction errors from the hypothalamus and other brainstem nuclei (Barrett & Simmons, 2015; Chanes & Barrett, 2016; Kleckner et al., 2017).

At the same time, several nodes of the default mode network and broader allostatic network did not show strong functional connectivity with the hypothalamus. Areas with weaker connectivity included the precuneus, most of the lateral temporal cortex, and the temporoparietal junction (including lateral inferior parietal lobe and angular gyrus), and also a pericingulate portion of the anterior prefrontal cortex (see **Figures 2** & **3**). Intriguingly, recent post-mortem studies have proposed that all of these areas have a distinct cytoarchitectural and structural connectivity pattern from the more anterior prefrontal nodes (Paquola et al., 2025). The posterior nodes and the perigenual prefrontal cortex have a gradient laminar pattern within each node, and are more closely connected with primary exteroceptive sensory inputs. In contrast, the prefrontal nodes of the default mode network have a more interdigitated laminar pattern within each node, and are more isolated from primary exteroceptive sensory inputs (visual and auditory). While the implications for these divisions for information processing and cognition remain to be seen, our functional connectivity findings provide further support for this model of differentiation within the default mode network.

Why might the anterior hypothalamus in particular show greater functional connectivity with these prefrontal nodes of the default mode network? The anterior region of the hypothalamus contains the preoptic and anterior hypothalamic nuclei. These nuclei play an important role in thermoregulation via metabolic (e.g., metabolizing fat for heat), autonomic (secretion of sweat), and goal-directed or behavioral (e.g., seeking shade) strategies (Morrison & Nakamura, 2011, 2019; Mota-Rojas et al., 2021; Siemens & Kamm, 2018). Thermoregulation is one of the most basic and important allostatic and metabolic challenges that an organism faces. Organisms can only operate optimally within a relatively narrow range of temperatures (Angilletta et al., 2010; Huey & Stevenson, 1979). Thus the need to monitor and regulate body temperature is a continuous process. One speculative possibility is that this relatively chronic need for thermoregulation may explain the stronger functional connectivity of the anterior hypothalamus with prefrontal, cingulate, and insular portions of the allostatic network, which themselves are known to be involved in goal-directed behavior and viscerosensory regulation.

**Conclusion**

We used high-resolution, high-field strength brain imaging to provide the first, data-driven functional parcellation of the human hypothalamus using resting state data. The results justified four functional divisions of the human hypothalamus. These divisions showed differential functional connectivity with two clusters of cortical areas. The anterior, followed by the superior, divisions of the hypothalamus exhibited greater functional connectivity particularly with certain prefrontal and limbic/paralimbic cortical areas includes nodes of the default mode network (e.g., anteromedial and ventrolateral PFC) and limbic/paralimbic regions (insula, cingulate cortex, anterior temporal cortex) that coalesce into an allostatic network - or a large scale network that supports the predictive regulation of the body state. These functional parcellations of the hypothalamus and their relations to cortical network architectures may be useful for guiding future work on understanding how the hypothalamus, and its functional-anatomic complexity, support a wide range of cognitive and behavioral phenomena for integrating the mind, brain, and body.

## Methods

### Participants

Adult, right-handed participants from the greater Boston area were recruited for the study (*N* = 104). Participants' ages ranged from 18 to 40 years (M = 26.85 , SD = 6.00). Of these participants, 60 identified as male and 44 identified as female. Eleven participants identified as Hispanic or Latino, 92 identified as Non Hispanic or Latino. 59 participants identified as White or Caucasian, 13 identified as Black or African-American, 29 identified as Asian and 2 chose not to answer. For educational level: 8 had completed graduate school, 18 had completed some graduate school, 26 had completed college, 38 completed some college, 11 completed high school or a GED, and 2 had completed some high school. Ethnicity, racial, and education data were missing for 1 participant due to a technical failure in the online data collection. Of the participants, 91 had all three runs, nine participants had only two runs, and four participants had only one run. In total, the analyze sample included 295 runs.

### fMRI Acquisition

Gradient-echo echo-planar imaging BOLD-fMRI was performed on a 7 Tesla MRI scanner at the Athinoula A. Martinos Center for Biomedical Imaging at Massachusetts General Hospital (MGH), Boston, MA. The scanner was built by Magnex Scientific (Oxford, UK), with the MRI console, gradient and gradient drivers, and patient table provided by Siemens. A custom-built 32-channel radiofrequency coil head array was used for reception. Radiofrequency transmission was provided by a detunable band-pass birdcage coil. Functional images were acquired using a GRAPPA-EPI sequence (GRAPPA acceleration factor = 3, TE = 28ms, TR = 2.34s, flip angle = 75°, 123 axial slices, A > P phase encoding, partial Fourier in the phase encode direction = 7/8). Structural images were also acquired using a GRAPPA-EPI sequence (GRAPPA acceleration factor = 3, TE = 22 ms, TR = 8.52 s, flip angle = 90°, 126 axial slices, A > P phase encoding, partial Fourier in the phase encode direction = 6/8). This structural EPI image was reconstructed (via freesurfer and custom scripts) into a T1-like image, which improved anatomical-functional registration and reduced blurring of functional signals by ensuring that anatomical and functional images had similar spatial distortions (Renvall et al., 2016). In both structural and functional images, voxels were 1.1mm isotropic (0mm gap between slices, FOV = 205 x 205 $mm^2$), echo spacing was 0.81ms, and bandwidth was 1415 Hz per pixel.

**fMRI Preprocessing**

Preprocessing of the anatomical and functional data was performed using the fmriprep pipeline, version 1.1.2 [1, 2, RRID:SCR_016216], a Nipype-based tool [3, 4, RRID:SCR_002502]. Pipeline details can be found at https://fmriprep.org/en/1.1.2/workflows.html. Each T1w (T1-weighted) volume was corrected for INU (intensity non-uniformity) using N4BiasFieldCorrection v2.1.0. Subject brain masks were computed by dilating a binary image of their skull-stripped T1 image by 2 voxels to remove gaps in coverage. Spatial normalization to the 2009c ICBM 152 Nonlinear Asymmetrical template was performed through nonlinear registration with the antsRegistration tool of ANTs v2.1.0, using brain-extracted versions of both T1w volume and template. Brain tissue segmentation of cerebrospinal fluid (CSF), white-matter (WM) and gray-matter (GM) was performed on the brain-extracted T1w using FSL fast (FSL v5.0.9). Functional data were slice time corrected using 3dTshift from AFNI v16.2.07 and motion corrected using FSL mcflirt. Mean framewise displacement was $M$ = .17, $SD$ = .07, interquartile range = .10. This was followed by co-registration to the corresponding T1w using boundary-based registration with 9 degrees of freedom, using FSL flirt. Motion correcting transformations, BOLD-to-T1w transformation and T1w-to-template (MNI) warp were concatenated and applied in a single step using antsApplyTransforms (ANTs v2.1.0) using Lanczos interpolation. Physiological noise regressors were extracted using the aCompCor method (Muschelli et al., 2014), taking the top five principal components from subject-specific CSF and WM masks, where the masks were generated by thresholding the WM/CSF masks derived from fast at 99% probability, constraining the CSF mask to the ventricles (using the ALVIN mask; (Kempton et al., 2011)), and eroding the WM mask using the binary erosion function in (SciPy v.1.6.1; Virtanen et al., 2020). Many internal operations of fmriprep use Nilearn, principally within the BOLD-processing workflow. For all participants, the quality of brain masks, tissue segmentation, and MNI registration was visually inspected for errors using the html figures provided by the fmriprep pipeline.

**General Linear Model**

In each subject, preprocessed functional BOLD data was submitted to a first-level GLM (FEAT; Woolrich et al., 2009; as implemented in nipype 1.1.4.dev0; Esteban et al., 2020) to remove noise attributable to motion or physiological artifacts. Nuisance regressors included 6 translation/rotation parameters, their temporal derivatives, their squares, and their squared temporal derivatives (Satterthwaite et al., 2013). Nuisance regressors also included 10 aCompCor terms, modeling the top 5 principal components of signal within whole-brain white matter and CSF masks (Muschelli et al., 2014), non-steady-state outliers (identified by fmriprep), and intercept, and a discrete cosine filter with  a 120 sec cutoff (simulating the effects of a high-pass filter within the GLM). No smoothing was performed, as all data would later be binned into ROI parcellations.

**Parcellations**

**Hypothalamic.** The probabilistic Pauli hypothalamus mask (in 2009c MNI space) was registered and resliced to native space voxel resolution (1.1 mm isotropic; nilearn.image.resample_img) and binarized at 20%, resulting in a mask with 1019 voxels. We masked out high variability voxels indicative of the third ventricle at the run level by sorting voxels based on signal variability over time and using a "kneedle" algorithm (Satopaa et al., 2011) to obtain a threshold value. In some cases, a voxel might be masked out due to excessive variability using the kneedle algorithm in one functional run, but not in the other ones, for a given subject's data. In those cases, we used the availability data for that voxel for analysis.

**Cortical**. We used the Glasser atlas (Glasser et al., 2016) for parcellation of cortical areas in volumetric space. The atlas includes 360 areas of cortical parcels (180 per hemisphere). We spatially normalized the Glasser parcels to our dataset using nilearn (Abraham et al., 2014)**.** For each subject, we concatenated the run-level data across three functional runs (i.e., residuals from the GLM), and then calculated a voxel-by-voxel (i.e. 1019x1019) functional connectivity matrix using the Pearson correlation coefficient. Because we masked out voxels with high variability on a run by run basis, there was the possibility that some voxels were masked out in some runs, but not others. In those cases, we used data from the available runs for analysis. We then Fisher transformed and averaged the participant-level functional connectivity matrices to estimate the group level functional connectivity matrix.

**Functional Connectivity Analysis**

Time series data per ROI were concatenated across runs per subject and intercorrelated, and the Fisher z-transformed, to generate a 364x364 (360 cortical ROIs and 4 hypothalamic subregion ROIs) functional connectivity matrix per subject. To reduce multiple comparison concerns and simplify results, we performed a k-means cluster analysis to group together ROIs with similar connectivity profiles across hypothalamic subregions (i.e. across a 360 cortical ROIs x 4 hypothalamic subregion ROIs, matrix). To identify the number of clusters that were justified by the data, we iteratively calculated the Calinski-Harabasz index and selected the maximum index (see **Supplemental Figure 1**). Fisher z-values for ROIs in the same k-means cluster were averaged together for each subject.

**Acknowledgements**

This research was supported by grants from the Brain and Cognitive Sciences Division of the National Science Foundation (award number: 2241938), the National Institutes of Health (award numbers: NCI U01 CA193632, R01 MH109464, R21 MH129902, F32 MH133288, and F32 MH122062), the National Institute on Aging (award number: NIA R01AG071173), and the Unlikely Collaborators Foundation. The views, opinions, and/or findings contained in this review are those of the authors and shall not be construed as necessarily reflecting the views of the Unlikely Collaborator Foundation

**Tables**

| Hypothalamic Subregion | *N* voxels | Mean Probability [25th P, 75th P] |
|---|---|---|
| Anteroventral-Tuberal Hypothalamus | 295 | .86 [.82,.97] |
| Anterior Hypothalamus | 234 | .91 [.88, 1.00] |
| Middle Tuberal-Posterior Hypothalamus | 291 | .74 [.63, .88] |
| Superior Hypothalamus | 199 | .88 [.74, .99] |

**Table 1**. Robustness of voxel assignment to hypothalamic subregion. **Table 1** shows the mean probability of a given pair of voxels being assigned to the same (vs. a different) subregion across 100 iterations of Louvain Community Detection analysis. Also shown are the 25th and 75th percentiles of those probabilities in brackets and the number of voxels within each hypothalamic subregion. P = Percentile.

## Figures

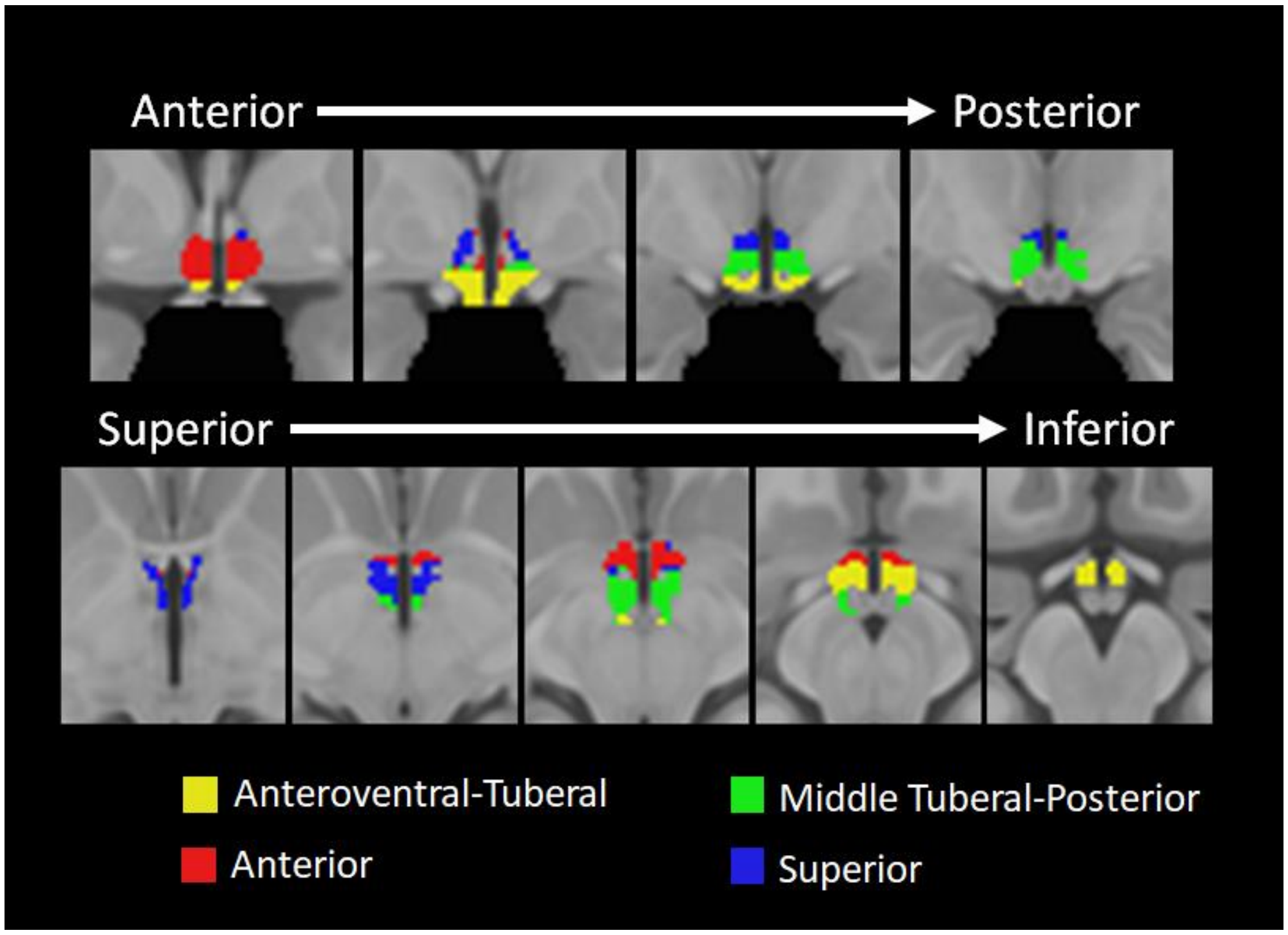

**Figure 1.** The four hypothalamic subregions identified using community detection analysis and two cortical communities identified with k-means clustering analysis. **Figure 1**, top row shows the coronal view of the hypothalamic subregions and moves from anterior to posterior (left to right). **Figure 1**, bottom row shows the axial view of the hypothalamic communities moving from superior to inferior (left to right). In the figure, the anteroventral-tuberal community is depicted in yellow, the anterior community is depicted in red, the middle tuberal-posterior community is depicted in green, and the superior community is depicted in blue.

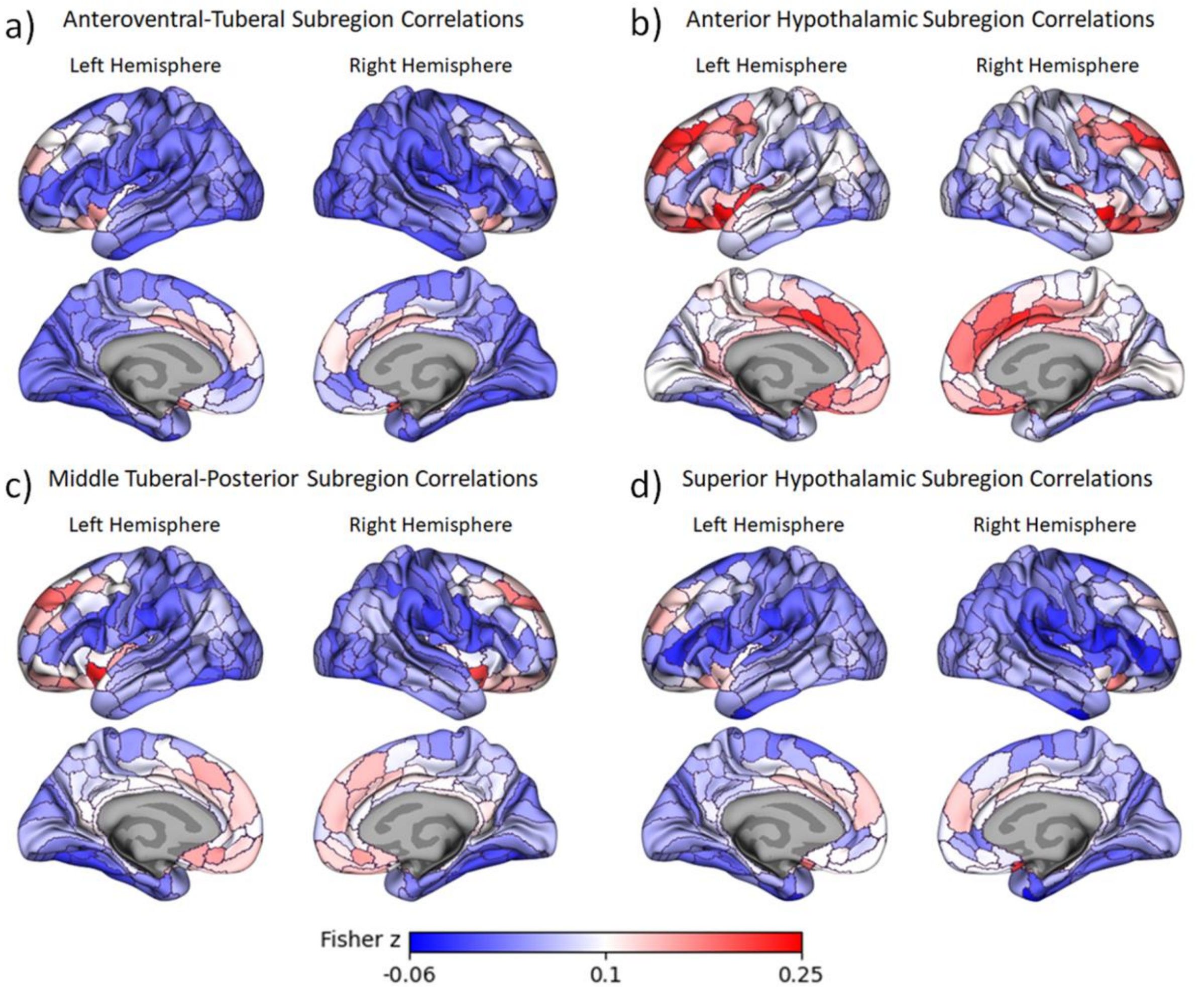

**Figure 2**. Correlations in Fisher's z between each hypothalamic subregion and each cortical parcel. Correlations for the anteroventral-tuberal subregion are shown in **2a**, correlations for the anterior subregion are shown in **2b**, correlations for the middle tuberal-posterior subregion are in **2c**, and correlations for the superior subregion are depicted in **2d**. Red depicts stronger positive correlations while blue depicts stronger negative correlations. In each figure, the left hemisphere is shown on the left and the right hemisphere is shown on the right. The lateral views are on the top row, while the medial views are on the bottom row.

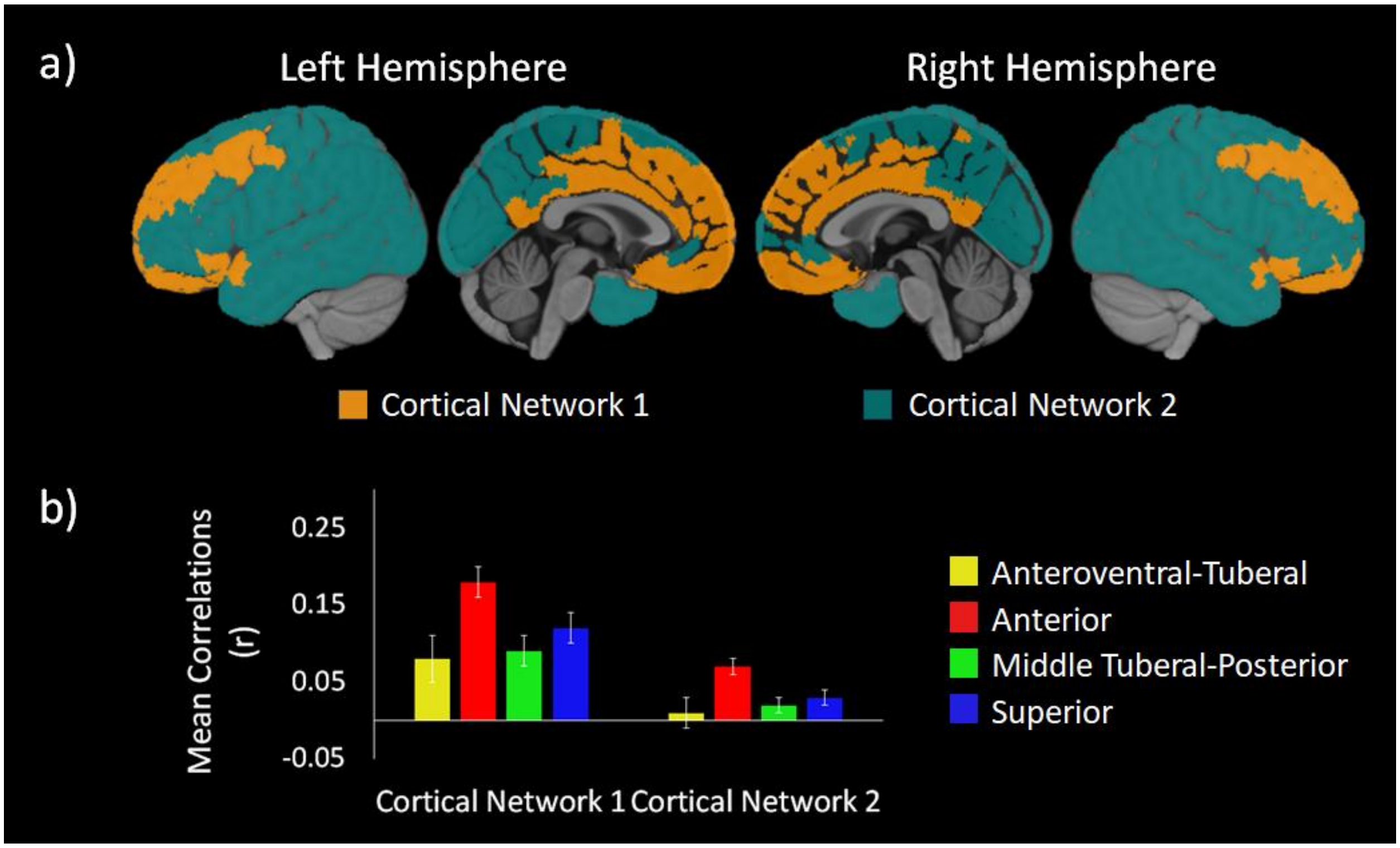

**Figure 3** shows the two network functional connectivity map and correlations between the cortical parcels and the hypothalamus. **Figure 3a** shows the cortical networks with Cortical Network 1 is in orange and Cortical Network 2 is in teal. **Figure 3b** shows the mean correlations between each cortical network and each hypothalamic community in the two-network solution. Each cortical network is plotted on the x-axis, while each hypothalamic community is depicted in different colors. The mean correlations are on the y-axis. Error bars reflect 95% confidence intervals. Overall, the hypothalamic communities showed greater connectivity with frontal and midline regions in Cortical Network 2 compared to the more posterior and lateral sensorimotor areas in Cortical Network 1. Moreover, the anterior hypothalamic community (Red) showed the greatest connectivity with both cortical networks compared to all other hypothalamic communities.

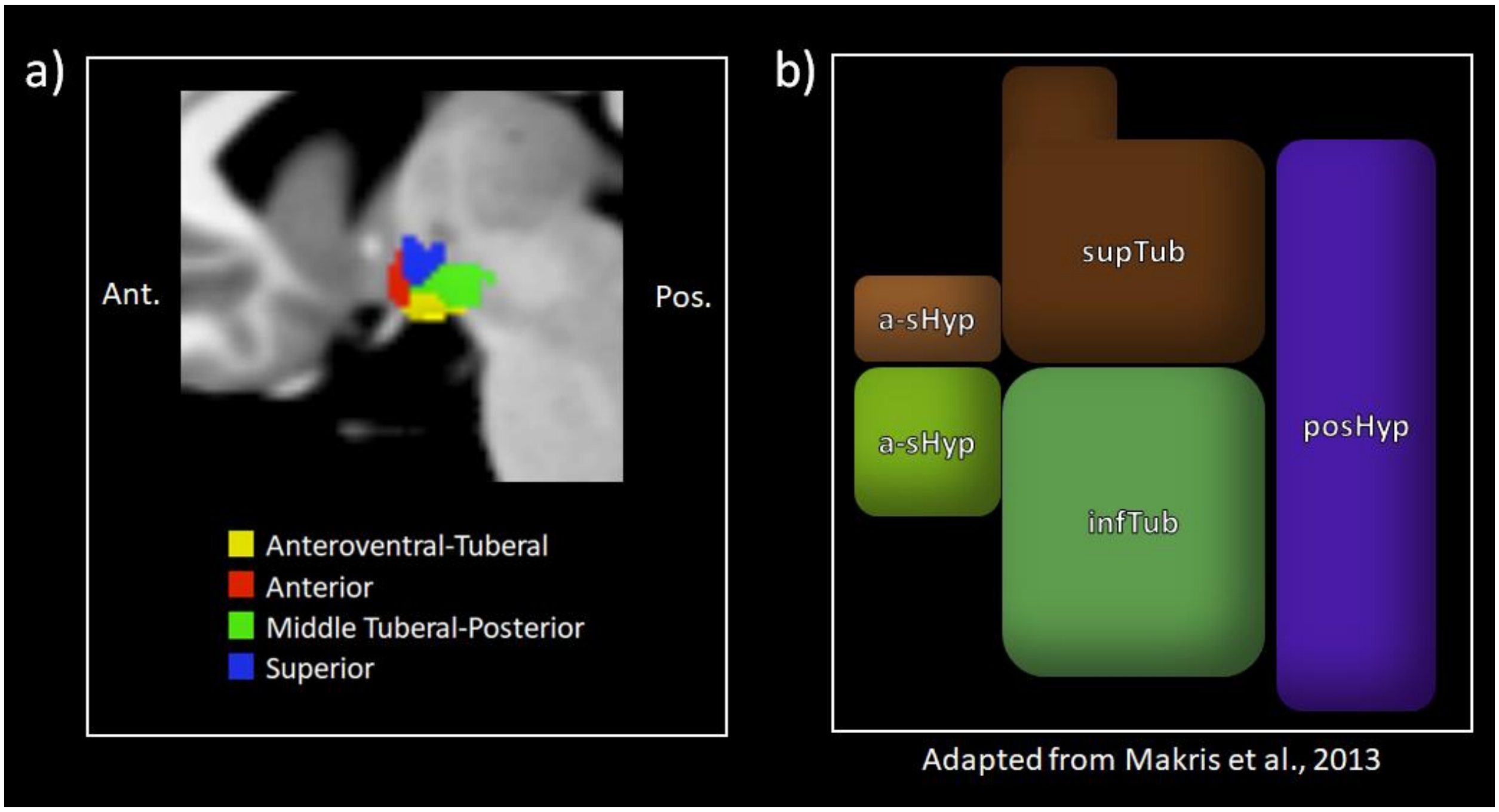

**Figure 4.** Comparison of functionally derived subregions in this study and anatomical parcels made by Makris et al. 2013. **4a** depicts the functional subregions identified in the present study in the medial view from the left hemisphere. and **4b** depicts Makris et al.'s (2013; FIgure 2) parcellation of the hypothalamus based on anatomical landmarks, adapted from their paper. Note that the third ventricle is visible in the figure from Makris et al. (2013) in tan in. Ant. = Anterior, Pos. = Posterior, a-sHyp = anterior-superior hypothalamus, a-iHyp = anterior-inferior hypothalamus, supTub = superior tuberal hypothalamus, infTub = inferior tuberal hypothalamus, posHyp = posterior hypothalamus

**Supplementary Figure**

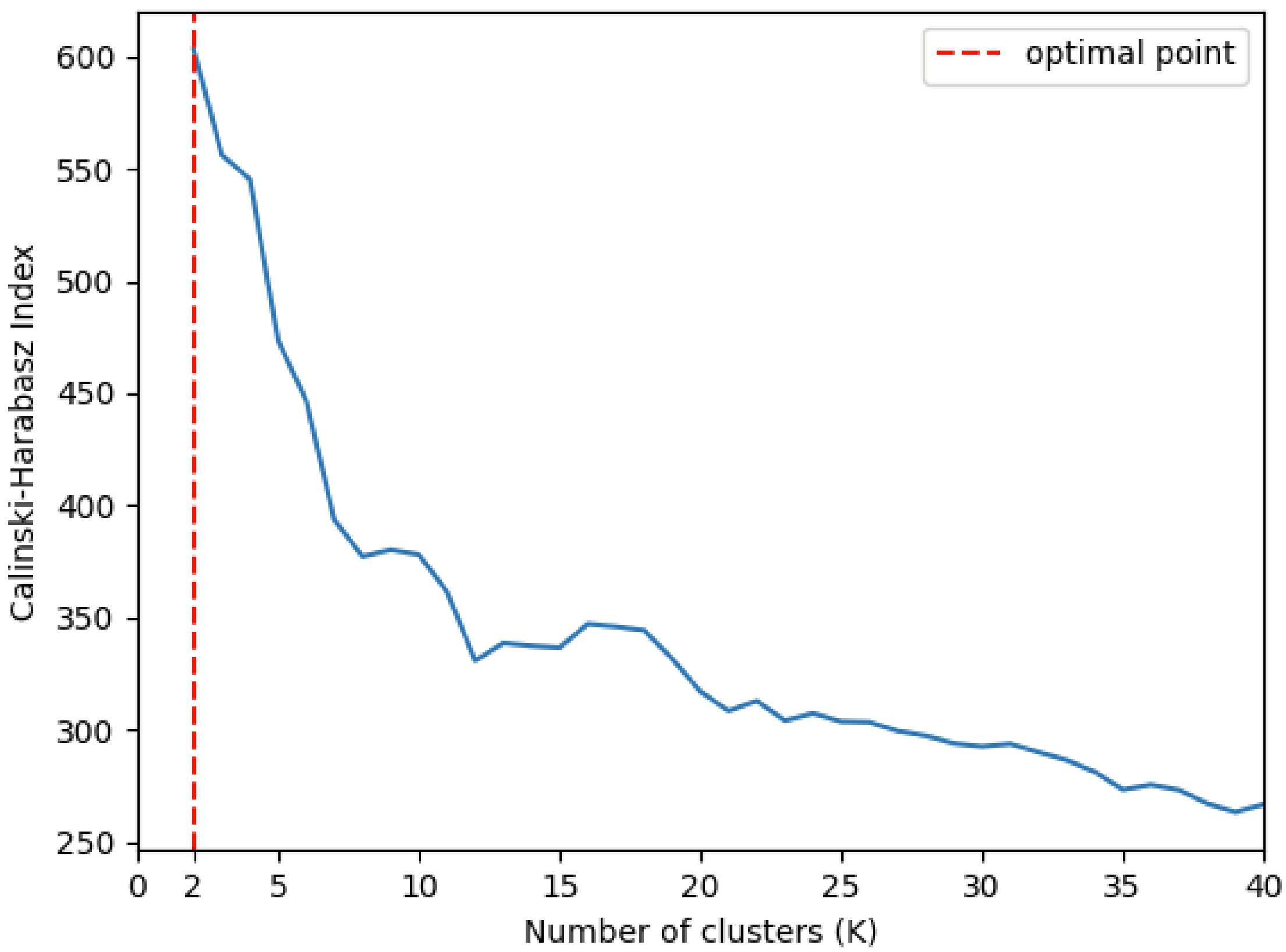

**Supplementary Figure 1**. Elbow plot of optimal number of clusters (Cortical Networks). The plot shows the Calinski-Harabasz index score on the Y-axis as the number of clusters (Cortical Networks) increases on the X-axis. Higher index scores reflect better clustering, as defined by greater between cluster variances vs. within cluster variance. The optimal number of clusters identified by the Kneedle algorithm was two, indicated by the red dashed line.